\documentclass[journal]{IEEEtran}
\usepackage{ifpdf}
\usepackage{url}
\ifpdf
\else
\fi
\usepackage{cite}
\ifCLASSINFOpdf
  \usepackage[pdftex]{graphicx}
  \graphicspath{{../pdf/}{../jpeg/}}
  \DeclareGraphicsExtensions{.pdf,.jpeg,.png}
\else
  \usepackage[dvips]{graphicx}
  \graphicspath{{../eps/}}
  \DeclareGraphicsExtensions{.eps}
\fi
\usepackage{amsmath}
\usepackage{algpseudocode}
\algtext*{EndWhile}
\algtext*{EndIf}
\algtext*{EndFor}
\usepackage{algorithm}
\usepackage{array}

\ifCLASSOPTIONcompsoc
 \usepackage[caption=false,font=normalsize,labelfont=sf,textfont=sf]{subfig}
\else
 \usepackage[caption=false,font=footnotesize]{subfig}
\fi
\usepackage{stfloats}
\usepackage{hyperref}
\usepackage{xcolor}
\long\def\comment#1{}

\newfont{\bbb}{msbm10 scaled 700}

\newfont{\bb}{msbm10 scaled 1100}

\newcommand{\av}{{\bf a}}

\newcommand{\dv}{{\bf d}}

\newcommand{\qv}{{\bf q}}
\newcommand{\rv}{{\bf r}}

\newcommand{\wv}{{\bf w}}

\newcommand{\xv}{{\bf x}}

\newcommand{\Dm}{{\bf D}}

\begin{document}

\title{A Derivation of Identifiable Condition for Non-Uniform Linear Array DOA Estimation}

\author{Hui Chen, Tarig Ballal, and Tareq Y. Al-Naffouri
\thanks{The authors are with the Division of Computer, Electrical and Mathematical Science \& Engineering, King Abdullah University of Science and Technology (KAUST), Thuwal, 23955-6900, KSA. e-mail: (\{hui.chen; tarig.ahmed; tareq.alnaffouri\}@kaust.edu.sa).}}

\maketitle

\begin{abstract}
Phase ambiguity happens in uniform linear arrays (ULAs) when the sensor distance is greater than $\lambda/2$. This problem in direction of arrival (DOA) estimation and can be solved by designing a proper sensor configuration. In this work, we derive the identifiable condition for ULA DOA estimation.
\end{abstract}


\IEEEpeerreviewmaketitle

\section{Observation Model}
\label{sec:model}
We consider a complex sinusoidal source signal, with a frequency $f$ and amplitude $A$, $s(t)=Ae^{-j2\pi f t}$ in the \emph{far field}~\cite{12-farfield} of a non-uniform linear array of $N$ sensors. The source impinges on the array from a direction $\theta_0 \in (-\pi/2, \pi/2)$ rad. By using $d_{uv}$ to denote the distance between two sensors ($u$ and $v$) normalized by $\lambda/2$, where $\lambda$ is the signal wavelength, the received signal (vector) at time $t$ can be modelled as~\cite{doa_model_coprime}
\begin{equation}
\xv(t) = \av(\theta_0) s(t) + \wv(t),
\label{eq_2}
\end{equation}
where $\av(\theta_0) = [1, \ \ e^{-j\pi d_{12}\text{sin}(\theta_0)},\ ..., \ e^{-j\pi d_{1N}\text{sin}(\theta_0)}]^T$ is the array steering vector, and $\wv(t)$ is vector of the additive noise.

The \emph{principal} phase difference across a sensor pair, $u$ and $v$, can be estimated from the $u$-th and $v$-th elements of $\xv$ as
\begin{equation}
\hat\psi_{uv}(t) = \mathrm{angle}(x_u(t) \cdot x_v^{*}(t)  ) \in[-\pi, \pi),
\label{phase_estimation}
\end{equation}
where $(\cdot)^*$ is the complex conjugate operation. Without loss of generality, we will focus on single-snapshot scenarios. Hence, we will subsequently drop the time variable $t$.

To develop our proposed method, we start from noise-free principal phase observations, $\psi_{uv}$. These observations are related to the actual phase difference, $\phi_{uv} = \pi d_{uv} \sin(\theta_0)$ as
\begin{equation}
\psi_{uv} = \mathrm{mod}({\phi}_{uv}+\pi,2\pi) - \pi = \pi d_{uv} \sin(\theta_0) - 2\pi q_{uv} ,
\label{eq_1}
\end{equation}
where $\mathrm{mod}(\cdot, \cdot)$ is the modulus operation, and $q_{uv}$ is an integer value given by the rounding operation
\begin{equation}
q_{uv} = \mathrm{round}\left(\frac{ \pi d_{uv} \sin(\theta_0) }{2\pi}\right).
\label{wrapped PD}
\end{equation}

Based on \eqref{eq_1}, we observe that estimating the DOA from $\psi_{uv}$ requires knowledge of the integer $q_{uv}$, which may not be available if a methods such as \eqref{phase_estimation} is used to estimate $\psi_{uv}$. For $d_{uv} \leq 1$, $q_{uv} = 0$ for any $\theta$. For $d_{uv} > 1$, the latter result is not guaranteed, except for a specific range of $\theta$. Since $\theta$ is unknown, $\psi_{uv}$ will always be \emph{ambiguous} for $d_{uv} > 1$.

\section{Identifiable Condition}
The concept of wrapped phase-difference pattern (WPDP) is introduced in~\cite{pdp_eusipco2019} to visualize phase-difference and estimate DOA.

From the WPDP, we can see that the sufficient and necessary condition is that there do not exist two DOAs that have the same WPD vectors. It is obvious that if two points have the same WPD vector, they cannot be differenciated from each other. Thus, we can have
\begin{equation}
\pi \text{sin}(\theta_1)\dv \neq \pi \text{sin}(\theta_2)\dv+2\pi \qv
\label{condition1}
\end{equation}
where $\theta_1>\theta_2$, $\qv = [q_1, q_2, ..., q_M]$ is a nonnegative integer vector indicating the possible wrapping cycle for each sensor pair. Define $q_{i,max}$ as the maximum integer that $q_i$ might be, $q_{i,max}$ can be calculated as
\begin{equation}
q_{i,max} = \mathrm{floor}\left(\frac{\pi (\sin(\theta_1)-\sin(\theta_2)) d_i}{2\pi}\right) \leq \mathrm{floor}(d_i).
\label{eq_tdoa_3}
\end{equation}

There are two situations:
\begin{enumerate}
    \item If there is no phase-wrapping in any sensor pair $i$, the inequation \eqref{condition1} holds because $q_i= q_{i,max}=0$ and $\theta_1 \ne \theta_2$. This is the case that the distance between one of the sensor pair is smaller than half-wavelength.
    \item If phase-wrapping happens for all the sensor pair, $q_i$ can be an integer from set $\{1, 2,..., q_{i,max}\}$. 
    Then, \eqref{condition1} can be reformulated as equation~\eqref{condition_nohold} {\bf{does not hold}} for all the possible value of $q_i$.
    \begin{equation}
    \frac{d_1}{q_1} = \frac{d_2}{q_2} = ... = \frac{d_M}{q_M} \left(= \frac{2}{\mathrm{sin}(\theta_1) - \mathrm{sin}(\theta_2)}\geq 1 \right).
    \label{condition_nohold}
    \end{equation}
    Note that $q_{i,max} \le \mathrm{floor}(d_i)$, the content inside the bracelet can be ignored.
\end{enumerate} 

Let us take two examples:

(a). An unidentifiable case with $\Delta=1.2, \delta=4$ provided by the reviewer. 

In this case, $\rv = [0, 1.2, 6]$, $\dv = [d_{12}, d_{13}, d_{23}] = [1.2, 6, 4.8]$, and $\qv_{max} = [1, 6, 4]$. if integer vector $\qv$ is chosen as $[1, 5, 4]$, equation~\eqref{condition_nohold} holds and hence it is an unidentifiable case.

(b). An identifiable case with $\Delta=3.6, \delta=1.25$ provided in Fig. 2.(a). 

In this case, $\rv = [0, 3.6, 8.1]$, $\dv = [d_{12}, d_{13}, d_{23}] = [3.6, 8.1, 4.5]$, and $\qv_{max} = [3, 8, 4]$. Whatever we choose the integer vector $\qv$, equation~\eqref{condition_nohold} cannot be satisfied and hence it is an identifiable case.
\\

\section{Quick Check of the Identifiability}
There is a quick way to check the condition in~\eqref{condition_nohold} is satisfied or not for a certain layout.
\begin{enumerate}
    \item Find a positive real number $I$, which makes $D_i=Id_i$ an integer for all the $i\in(1,2,...,M)$ and the greatest common divisor for ${D_1, D_2,... D_M}$ is $1$; 
    \item Since $q_i$ is an integer and ${D_1, D_2,... D_M}$ have the greatest common divisor $1$, the only way to make $\frac{D_1}{q_1} = \frac{D_2}{q_2} = ... = \frac{D_M}{q_M}$ is to choose $q_i$ equals to $D_i$ or equals to multiple times of $D_i$. 
    \item If $D_i \le p_{i,max}$ for $i = 1,2,...,M$, equation~\eqref{condition_nohold} is satisfied.
\end{enumerate}

Let us take the same two examples:

(a). An unidentifiable case with $\Delta=1.2, \delta=4$ provided by the reviewer. 

Multiply $\dv$ by 10 to obtain $[12, 60, 48]$, then divided by the greatest common divisor to obtain $\Dm = [1, 5, 4]$ ($I = \frac{5}{6}$). Because $D_i \le q_{i,max}$ for $i = 1,2,...,M$, equation~\eqref{condition_nohold} is satisfied and hence this configuration is unidentifiable.

(b). An identifiable case with $\Delta=3.6, \delta=1.25$ provided in Fig. 2.(a). 

Multiply  $\dv$ by 10 to obtain $[36, 81, 45]$, then divided by the greatest common divisor to obtain $\Dm = [4, 9, 5]$ ($I = \frac{10}{9}$). Because $D_i > q_{i,max}$ for some $i$, equation~\eqref{condition_nohold} is not satisfied and hence this configuration is identifiable.

\section{Conclusion}
In this work, we briefly described the DOA estimation model in a far field scenario. An identifiable condition is derived based on the wrapped phase-difference pattern (WPDP), and a quick check approach is provided.

\ifCLASSOPTIONcaptionsoff
\newpage
\fi

\bibliographystyle{IEEEtran}
\bibliography{refs}

\begin{thebibliography}{1}
\providecommand{\url}[1]{#1}
\csname url@samestyle\endcsname
\providecommand{\newblock}{\relax}
\providecommand{\bibinfo}[2]{#2}
\providecommand{\BIBentrySTDinterwordspacing}{\spaceskip=0pt\relax}
\providecommand{\BIBentryALTinterwordstretchfactor}{4}
\providecommand{\BIBentryALTinterwordspacing}{\spaceskip=\fontdimen2\font plus
\BIBentryALTinterwordstretchfactor\fontdimen3\font minus
  \fontdimen4\font\relax}
\providecommand{\BIBforeignlanguage}[2]{{%
\expandafter\ifx\csname l@#1\endcsname\relax
\typeout{** WARNING: IEEEtran.bst: No hyphenation pattern has been}%
\typeout{** loaded for the language `#1'. Using the pattern for}%
\typeout{** the default language instead.}%
\else
\language=\csname l@#1\endcsname
\fi
#2}}
\providecommand{\BIBdecl}{\relax}
\BIBdecl

\bibitem{12-farfield}
J.~R. Gonzalez and C.~J. Bleakley, ``High-precision robust broadband ultrasonic
  location and orientation estimation,'' \emph{IEEE Journal of selected topics
  in Signal Processing}, vol.~3, no.~5, pp. 832--844, 2009.

\bibitem{doa_model_coprime}
C.~Zhou, Y.~Gu, X.~Fan, Z.~Shi, G.~Mao, and Y.~D. Zhang, ``Direction-of-arrival
  estimation for coprime array via virtual array interpolation,'' \emph{IEEE
  Transactions on Signal Processing}, vol.~66, no.~22, pp. 5956--5971, 2018.

\bibitem{pdp_eusipco2019}
H.~Chen, T.~Ballal, X.~Liu, and T.~Y. Al-Naffouri, ``Realtime 2-d doa
  estimation using phase-difference projection (pdp),'' in \emph{2019 27th
  European Signal Processing Conference (EUSIPCO)}.\hskip 1em plus 0.5em minus
  0.4em\relax IEEE, 2019, pp. 1--5.

\end{thebibliography}

\end{document}